\documentclass[%
 reprint,
 amsmath,amssymb,
 aps,
]{revtex4-1}

\usepackage{graphicx}% Include figure files
\usepackage{dcolumn}% Align table columns on decimal point
\usepackage{bm}% bold math
\begin{document}

\preprint{APS/PRL}

\title{Photon Blockade in Two-Emitter-Cavity Systems }% Force line breaks with \\

\author{Marina Radulaski}
 \email{marina.radulaski@stanford.edu}
\author{Kevin A. Fischer}
\author{Konstantinos G. Lagoudakis}
\author{Jingyuan Linda Zhang}
\author{Jelena Vu\v ckovi\' c}

\affiliation{%
E. L. Ginzton Laboratory, Stanford University, Stanford, California 94305, United States
}%

\date{\today}% It is always \today, today,

\begin{abstract}
The photon blockade (PB) effect in emitter-cavity systems depends on the anharmonicity of the ladder of dressed energy eigenstates. The recent developments in color center photonics are leading toward experimental demonstrations of multi-emitter-cavity solid-state systems with an expanded set of energy levels compared to the traditionally studied single-emitter systems. We focus on the case of $N = 2$ nonidentical quasi-atoms strongly coupled to a nanocavity in the bad cavity regime (with parameters within reach of the color center systems), and discover three PB mechanisms: polaritonic, subradiant and unconventional. The polaritonic PB, which is the conventional mechanism studied in single-emitter-cavity systems, also occurs at the polariton frequencies in multi-emitter systems. The subradiant PB is a new interference effect owing to the inhomogeneous broadening of the emitters which results in a purer and a more robust single photon emission than the polaritonic PB. The unconventional PB in the modeled system corresponds to the suppression of the single- and two-photon correlation statistics and the enhancement of the three-photon correlation statistic. Using the effective Hamiltonian approach, we unravel the origin and the time-domain evolution of these phenomena.
\end{abstract}

\maketitle

\emph{Introduction}---While an arbitrary number of photons can populate a bare nanocavity, the paradigm changes with the introduction of a strongly coupled dipole emitter. The photon blockade (PB) effect prevents the absorption of the second photon at specific frequencies due to the nonlinearity of the emitter that dresses the energy states and leads to an anharmonic ladder. This effect has been extensively studied in single-emitter ($N=1$) atomic \cite{atomic, birnbaum2005photon} and quantum dot \cite{detuned, dietrich2016gaas} cavity quantum electrodynamics (CQED), as well as in circuit QED systems \cite{lang2011observation}. Here, the effect occurs at the frequencies of the dressed states so-called polaritons (polaritonic PB) and results in a faster emission rate of single photons compared to the bare emitter. Experimentally, a signature of single photon emission has been the reduced value of the second order coherence $g^{(2)}(0)<1$. A recent paper has contested this criterion \cite{unconventional_blockade}, demonstrating conditions for the so-called \emph{unconventional} photon blockade where the two-photon statistic is suppressed, but the enhanced higher order coherences strengthen the generation of multiple photons. A cavity coupled to multiple ($N\gtrsim 1$) emitters would offer a richer set of dressed states and extend new opportunities for nonclassical light generation with applications in quantum key distribution \cite{o2009photonic}, quantum metrology \cite{giovannetti2006quantum} and quantum computation \cite{knill2001scheme}. Moreover, the collective coupling rate $G_N = \sqrt{\sum_{n=1}^{N}g_n^2}$ which sets the CQED device operating speed would effectively increase by a factor of $\sqrt{N}$ from the single-emitter case \cite{atomic_MECQED, andrei}.

To achieve the collective strong coupling of multiple emitters to a nanocavity, the collective coupling strength has to dominate over the cavity ($\kappa$) and emitter ($\gamma$) linewidth-induced loss mechanisms: $G_N>\kappa/4, \gamma/4$, as well as be greater than or comparable to the inhomogeneous broadening ($\Delta$) in the system: $\Delta \lesssim G_N$ \cite{inhomogeneousCQED}. This regime has been demonstrated in atomic systems \cite{two_atoms} but its implementation in solid-state would result in a three orders of magnitude speedup and a potential for on-chip integration. Among the solid-state emitters, color centers in diamond and silicon carbide \cite{SiV_broadening, Vsi_broadening} are the ones that feature high dipole moment and small inhomogeneous broadening ($\Delta < 30$ GHz) needed for fast and scalable nanophotonics platform. Moreover, their integration with nanocavities is a topic of active research \cite{burek, hybrid, hu, alp} paving the way for experimental demonstrations of color center based CQED.

\begin{figure}[t]
\includegraphics[width=78mm]{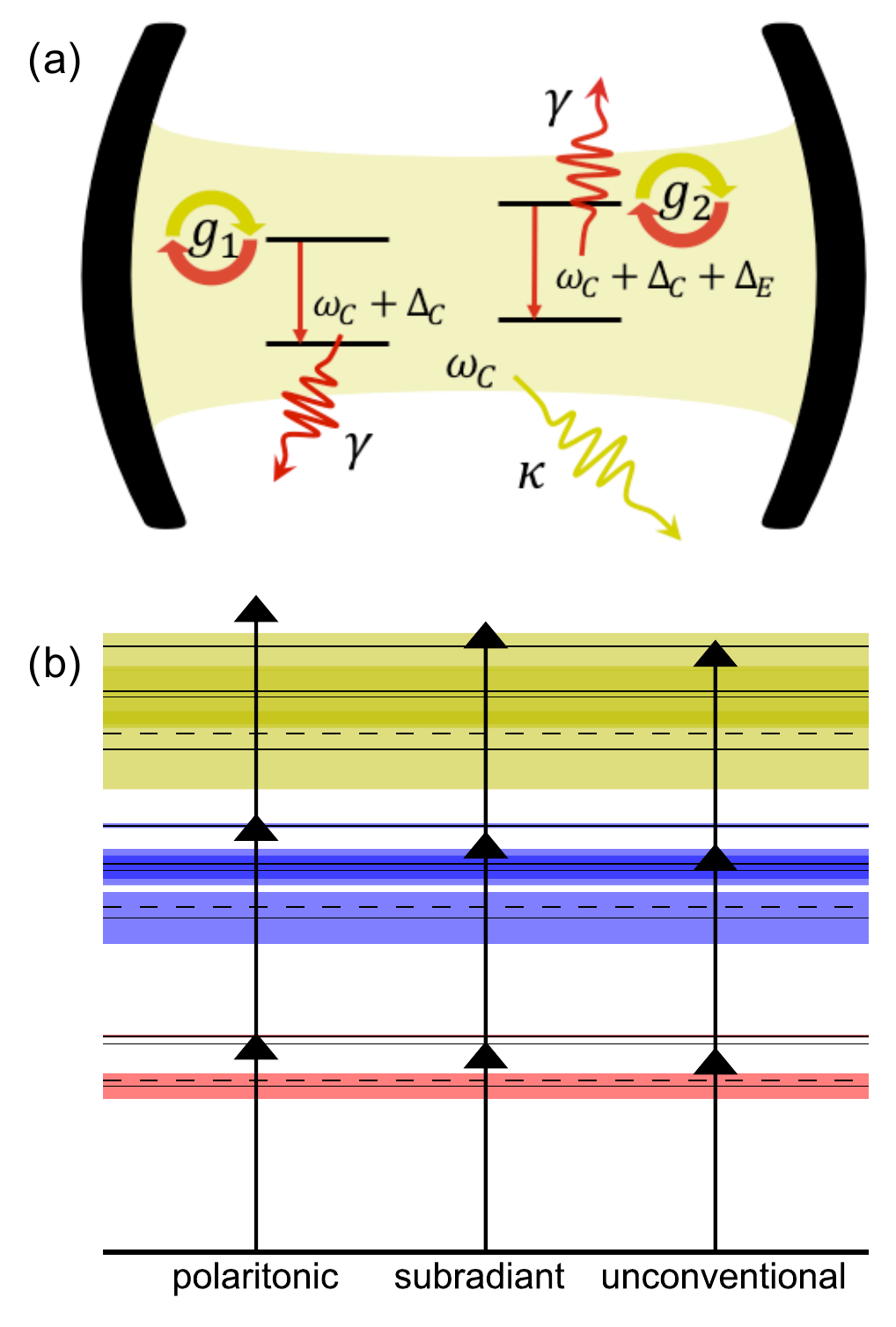}
\caption{\label{fig1} (a) An $N=2$ multi-emitter CQED system scheme and (b) an example of a dressed ladder with arrows illustrating photon processes in three types of photon blockade annotated at the bottom; solid (dashed) black lines represent energy levels of a dressed (bare) cavity, dressed levels are represented with linewidths; red, blue and yellow states belong to the first, second and third rungs, respectively; spacing between rungs is not to scale; $\{\kappa, \gamma, g_1, g_2, \Delta_C, \Delta_E\}/2\pi = \{25, 0.1, 10, 10, 30, 5\}$ GHz.}
\end{figure}

We consider theoretically the coherence effects in an $N=2$ multi-emitter CQED system [Fig. \ref{fig1}(a)] and analyze the influence of inhomogeneous broadening to the photon blockade effect. Using the quantum master equation with an extended Tavis-Cummings model, we discover conditions for polaritonic, subradiant and unconventional PB [Fig. \ref{fig1}(b)]. The subradiant PB is a new interference effect owing to the inhomogeneous broadening of the emitters which results in a purer and a more robust single photon emission than the polaritonic PB. Using the effective Hamiltonian approach, we describe the origin and the time-domain evolution of these phenomena.

\emph{The model}---The interaction Hamiltonian for our CQED system consists of the cavity, emitter and coupling terms ($\hbar = 1$):
\begin{equation}
H_I = \omega_C a^\dag a + \sum_{n=1}^N{[ \omega_{En} \sigma_n^\dag \sigma_n + g_n (\sigma_n^\dag a + a^\dag \sigma_n) ]},
\end{equation}
where $a$ and $\omega_C$ represent the annihilation operator and resonant frequency of the cavity mode; $\sigma_n$, $g_n$ and $\omega_{En}$ are the lowering operator, cavity coupling strength and transition frequency of the $n$-th out of $N = 2$ emitters. To treat the cavity and emitter detunings more explicitly, we rewrite emitter frequencies as $\omega_{E1} = \omega_C + \Delta_C$ and $\omega_{E2} = \omega_C + \Delta_C + \Delta_E$ and the Hamiltonian transforms into:
\begin{multline}
H_I^{N=2} = \omega_C a^\dag a +  (\omega_C + \Delta_C) \sigma_1^\dag \sigma_1 + (\omega_C + \Delta_C + \Delta_E) \sigma_2^\dag \sigma_2 \\
 + g_1 (\sigma_1^\dag a + a^\dag \sigma_1) + g_2 (\sigma_2^\dag a + a^\dag \sigma_2),
\end{multline}

\begin{figure*}[t]
\includegraphics[width=180mm]{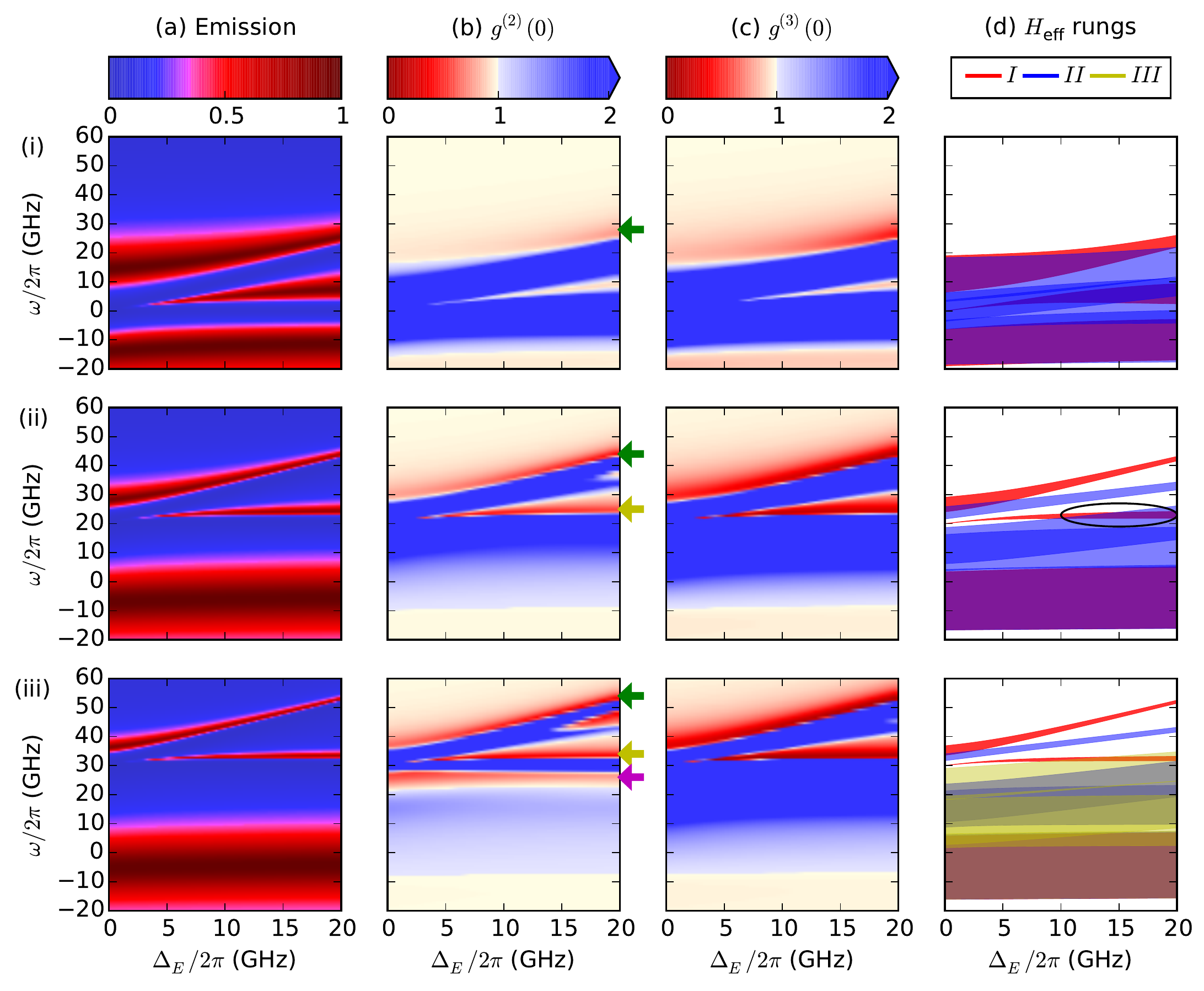}
\caption{\label{fig2} (a) Emission spectrum, (b) equal-time second- and (c) third-order coherence of transmitted light calculated by the quantum master equation. Green, yellow and magenta arrows indicate parameters of polaritonic, subradiant and unconventional PB, respectively. (d) Frequency overlap between the $k$-th order rungs of the dressed ladder of states of $H_{\textsf{eff}}$, presented as $\omega_{k}/k -\omega_C$ with indicated linewidths, first, second and third rung are marked by red, blue and yellow areas, respectively; $\Delta_C/2\pi$ takes values of (i) 0, (ii) 20 and (iii) 30 GHz. The circled area indicates a contradictory region of PB.}
\end{figure*}

The system is characterized by the cavity energy decay and emitter linewidth $\{\kappa, \gamma\}/2\pi = \{25, 0.1\}$ GHz, corresponding to quality factor $Q \approx 15,000$, emitter lifetime $\tau = 10$ ns, and the individual emitter-cavity coupling rate $g_n /2\pi = 10$ GHz. Coupling rate is defined by the emitter lifetime, emission branching ratio into the zero-phonon line $\rho_{ZPL}$, cavity index of refraction $n$, mode volume $V$, and the relative electric field the dipole experiences locally (electric field and dipole orientations form the angle $\phi$) in the resonant mode $\eta = \left| E(r)\cos \phi/E_{max} \right|$ as $g = \eta \sqrt{3\pi c^3 \rho_{ZPL}/2 \tau \omega_C^2 n^3 V}$ \cite{AAMOP}. Comparing to the state-of-the-art results with silicon-vacancy centers in a photonic crystal cavity \cite{alp}, the quoted values could be achieved by designing nanocavities with doubled quality factor and five-fold increased coupling rate, which can be achieved by reducing the mode volume several times and providing higher-precision positioning of the color centers at the field maximum \cite{schroder2016scalable}. The exact value of the emitter linewidth (here, three-fold smaller), which is the smallest system rate in the bad cavity regime, does not change the dynamics significantly when its order of magnitude is maintained.

To model the interaction of the system with its environment, we use the quantum master equation to calculate the steady state solution of the density matrix $\rho$:
\begin{multline}
\dot \rho = -i[H_I + E_P(a e^{\textrm{\footnotesize i}\omega t}+a^\dagger e^{-\textrm{\footnotesize i}\omega t}), \rho] +\kappa \mathcal{L}[a] \\
+ \sum_{n=1}^N\gamma \mathcal{L}[\sigma_n],
\end{multline}
where $E_P = \kappa/50$ represents the laser field amplitude, and loss terms are introduced through the superoperator $\mathcal{L}[O] = O\rho(t)O^\dag-\frac{1}{2}\rho(t)O^\dag O-\frac{1}{2}O^\dag O\rho(t).$ The system is then transformed into a rotating frame to remove the time-dependence \cite{steck} and obtain the steady state solution. When calculating emission spectra as $\langle a^\dagger a \rangle$, we leave the $E_P$ term out and add a pump term in the Liouvillian as $P \mathcal{L}[a^\dagger]$, where $P$ represents the laser power.

We gain a more intuitive understanding of the system dynamics by diagonalizing the effective Hamiltonian, constructed with the complex frequencies that account for cavity and emitter loss \cite{inhomogeneousCQED}:
\begin{equation}
H_{\textsf{eff}}=H_I-i \frac{\kappa}{2}a^\dagger a- i \sum_{n=1}^N \frac{\gamma}{2} \sigma^\dagger_n\sigma_n.
\end{equation}
The real and the imaginary part of its eigenvalues represent frequencies and half-linewidths of the excited energy states, respectively, while the eigenvectors quantify the cavity-like and the emitter-like character of the excited state. Figure \ref{fig1}(b) shows the dressed ladder of state for a sample system of nonidentical emitters detuned from the cavity. The energy levels are illustrated with their linewidths which form the absorption zones for $n$-photon events. In contrast to CQED systems with a single emitter where all excited states contain two levels, here we see that an additional emitter generates new excited states. The new state in the first rung resembles the wavefunction of a \emph{subradiant} state known from atomic systems not to couple to the environment well. There are also two new states in the second rung. These additional levels ultimately lead to a much richer set of physics phenomena. Not only do we find an enhanced regime of photon blockade from the subradiant states, but a newly discovered unconventional photon blockade regime \cite{unconventional_blockade} can also be achieved in the system. We now explore these effects as a function of the emitter detuning $\Delta_E$.

\emph{Results}---We present calculation results for cavity detunings $\Delta_C/2\pi=$ 0, 20 and 30 GHz which capture system's trends and features. The emission spectra are shown in Fig. \ref{fig2}(a). In parallel, we calculate the eigenstates of the first rung of the effective Hamiltonian, shown as red surfaces in Fig. \ref{fig2}(d). The three transmission peaks are in a close agreement with these eigenfrequencies, which expectedly indicates that the bottom peak is cavity-like, and the top two peaks emitter-like. To understand the effects that non-identical emitters bring into CQED more closely, we now focus on the emerging subradiant state for $\Delta_E/2\pi \leq 3$ GHz. Qualitatively similar spectra have been calculated for superconducting cavities coupled to an ensemble of spins \cite{inhomogeneous_peaks} and experimentally observed in superconducting circuits \cite{superconducting}. The collective strong coupling rate $G_2=\sqrt{g_{1}^2+g_{2}^2}$ places the two polariton peak frequencies close to $\omega_{pol\pm}^{\Delta_E = 0}=\omega_C + \frac{\Delta_C}{2}-i\frac{\kappa + \gamma}{4} \pm \frac{\sqrt{4\Delta_C^2+16G_2^2+ 2\kappa\gamma +4i \Delta_C \kappa - 4i\Delta_C\gamma + i \kappa^2 + i \gamma^2}}{4}$. Here, we describe the subradiant state in more detail by deriving approximations to its frequency $\omega_{sub}$ and state vector $v_{sub}$ for $0<\Delta_E \ll G_2, \frac{\kappa - \gamma}{2}$:
\begin{equation}
\omega_{sub} \approx \omega_C + \Delta_C + \frac{g_{1}^2}{G_2^2}\Delta_E - i \left(\frac{\gamma}{2} + \frac{\kappa - \gamma}{8G_2^2}\Delta_E^2\right),
\end{equation}
\begin{multline}
v_{sub} \approx \frac{1}{\sqrt{A}} \bigg\{ \left(-\frac{g_{2}\Delta_E}{G_2^2}+i\frac{\kappa - \gamma}{8g_{2}}\frac{\Delta_E^2}{G_2^2}\right) a^\dag \\
+ \left[- \frac{g_{1}}{g_{2}} \frac{8g_{2}^2+i(\kappa - \gamma)\Delta_E}{8g_{1}^2-i(\kappa - \gamma)\Delta_E} \right] \sigma_1^\dag + \sigma_2^\dag \bigg\} \left| 0 \right>,
\end{multline}
where $A$ is a normalization factor and $\left| 0 \right>$ represents the state with empty cavity and all emitters in the ground state. For vanishing $\Delta_E$ the cavity term in $v_{sub}$ becomes zero, diminishing state's coupling to the environment. With an increasing $\Delta_E$, the frequency of the subradiant state grows linearly, while the linewidth and the amplitude increase quadratically, closely matching the trend in the simulated emission spectra for $\Delta_E/2\pi \leq 3$ GHz.

Next, we quantify the system's photon blockade by analyzing the equal-time second- and third-order coherence of light transmitted through the system [Figs. \ref{fig2}(b-c)], $g^{(n)}(0)=\langle (a^\dag)^na^n \rangle/\langle a^\dag a\rangle^n$. Areas of suppressed $g^{(2)}(0)$ statistics (red areas) indicate the potential for single-photon emission, which is further supported if the $g^{(3)}(0)$ values are simultaneously reduced. Polaritonic PB is identified close to the frequencies of the emission peaks (green arrows in Fig. \ref{fig2}b), and its quality improves for higher detuned emitter-like peaks. This is consistent with the findings in single-emitter CQED systems where the photon blockade at the polariton peak frequency strengthens with the emitter-cavity detuning \cite{detuned}.

\begin{figure}[t]
\includegraphics[width=88mm]{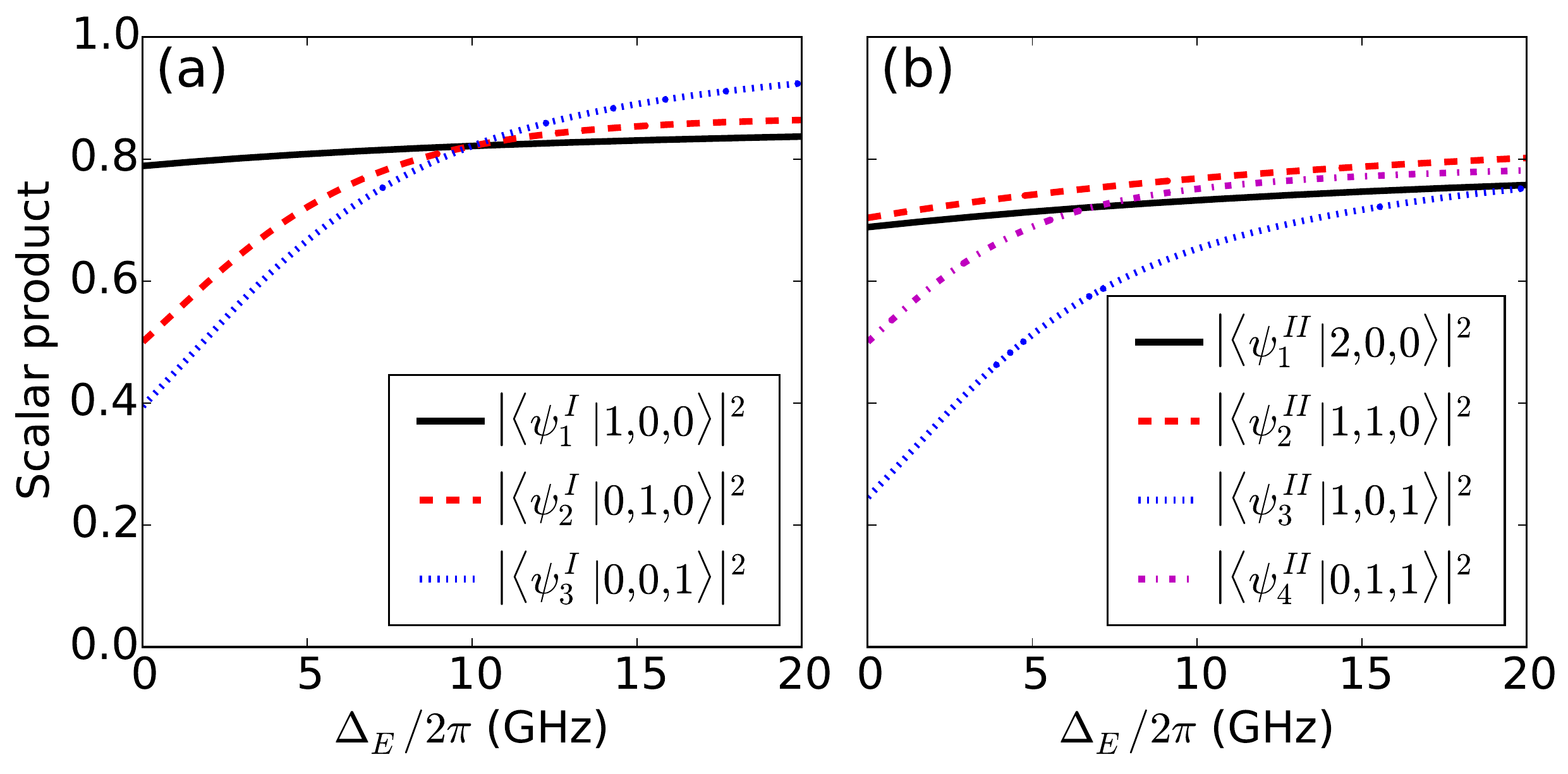}
\caption{\label{fig3} Scalar product of the dressed eigenstates of (a) the first and (b) the second rung of $H_{\textsf{eff}}$ and their main contributing bare states for $\Delta_C =$ 20 GHz.}
\end{figure}

We identify a novel effect of reduced second order coherence value at the frequency of the subradiant peak -- the \emph{subradiant} photon blockade (yellow arrows in Fig. \ref{fig2}b). While this effect has no direct analog in single emitter CQED we discover its origin with the help of the effective Hamiltonian approach. Analyzing the frequency overlap ${E_k/k-\omega_C}$ between eigenstates of different order ($k$) rungs of $H_{\textsf{eff}}$ [Fig. \ref{fig2}(d)] we find that the two photon emission is suppressed for the frequencies where single photon absorption is possible (red areas), but the second photon absorption is not (no overlapping blue areas which would, together with red ones, form violet areas in the plot). Surprisingly, for large $\Delta_E$ this condition disappears for the subradiant peak, but the subradiant photon blockade still persists (circled region in Fig. \ref{fig2}d). To understand this apparent contradiction, we perform an additional analysis of the eigenstate character for $\Delta_C/2\pi =$ 20 GHz, presented in Fig. \ref{fig3}. We compare dressed states to the bare states in $\left| C, E_1, E_2\right>$ basis, whose terms represent the cavity, the first and the second emitter excitations, respectively. We find that for an increasing $\Delta_E$ the eigenstates in the first two rungs of the ladder $\psi_1^I, \psi_2^I, \psi_3^I, \psi_1^{II}, \psi_2^{II}, \psi_3^{II}$ and $\psi_4^{II}$ (enumerated in an increasing energy order), have the highest scalar product with bare states $\left|1,0,0 \right>, \left| 0,1,0 \right>, \left| 0,0,1 \right>, \left|2,0,0 \right>, \left| 1,1,0 \right>, \left| 1,0,1 \right>$, and $\left| 0,1,1 \right>$, respectively. In other words, they start behaving like the corresponding bare states. Therefore, the dipolar coupling between the subradiant state $\psi_2^I$ and doubly excited state $\psi_3^{II}$ has to be inhibited due to $\left< 1,0,1 | a^{\dag}| 0,1,0\right>=0$, at higher detunings, which in turn suppresses the second photon absorption. With this combination of spectral and vector component properties obtained from the diagonalization of the effective Hamiltonian we predict parameters that give rise to enhanced single photon emission. The properties also hold for nonidentically coupled emitters ($g_{1} \neq g_{2}, G_2>\kappa/2$), which is favorable for systems with randomly positioned color centers whose coupling strength can not be imposed uniformly.

In addition to these trends, we also find conditions for realization of a superbunching effect recently dubbed the \emph{unconventional} photon blockade \cite{unconventional_blockade}, which confirms that the reduction in two-photon statistics is not a sufficient condition for single-photon emission. This effect occurs in the system with $\Delta_C/2\pi=30$ GHz in the region around $\omega/2\pi = 25$ GHz (magenta arrow in Fig. \ref{fig2}b) which has a low $g^{(2)}(0)$ but high $g^{(3)}(0)$ value, indicating preferential three-photon emission. To understand the occurrence of this regime in our system, we look into the third rung of the dressed ladder shown in yellow at Fig. \ref{fig2}(d-iii) and reveal that this frequency region has an overlapping three-photon absorbing process, but no two-photon absorbing process, which explains the calculated statistics. Thus, the multi-emitter CQED system will not just advance the single-photon generation, but also allow for the exploration of exciting regimes of multi-photon physics and statistics.

\begin{figure}[b]
\includegraphics[width=88mm]{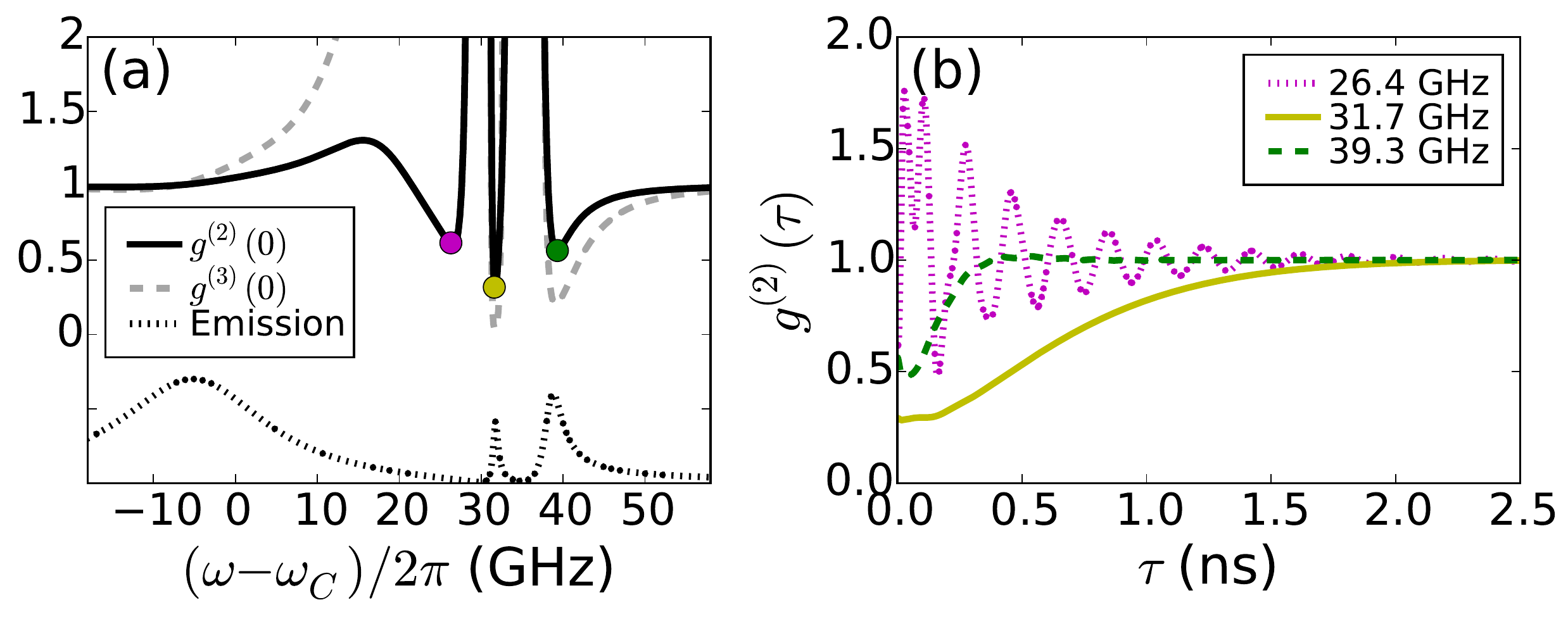}
\caption{\label{fig4} (a) Zero-delay second- and third-order coherences for $\{\Delta_C,\Delta_E\}/2\pi = \{30,5\}$ GHz system as a function of laser detuning from the cavity; emission spectrum in dashed lines illustrates the relationship between features. (b) $g^{(2)}(\tau)$ at frequencies of the corresponding $g^{(2)}(0)$ minima from the plot (a) representing unconventional (magenta/dotted), subradiant (yellow/solid) and polaritonic (green/dashed) photon blockade.}
\end{figure}

Finally, we analyze the system dynamics in the time domain in terms of interferences between the excited states. We look into the three $g^{(2)}(0)$ dips for $\{\Delta_C, \Delta_E\}/2\pi = \{30, 5\}$ GHz [Fig. \ref{fig4}(a)] which represent unconventional (magenta/dotted), subradiant (yellow/solid) and polaritonic (green/dashed) photon blockade, respectively. The second-order coherence evolution of the unconventional photon blockade frequency [magenta/dotted line in Fig. \ref{fig4}(b)] represents a damped oscillation and once again confirms that the system can not be a good single photon source at these frequencies. The dominant oscillation frequency ($\omega/2\pi$) is 5.2 GHz, while the first 200 ps also exhibit additional 11 GHz oscillation. The origin of the unconventional PB oscillations is currently unclear and will require more theoretical consideration in future. The subradiant and polaritonic PB second-order coherence traces represent a decay to an uncorrelated statistics. The characteristic half-width at half-maximum times are 0.7 ns and 0.2 ns, respectively, and represent a speedup in single photon emission compared to a bare quasi-emitter (10 ns). Both functions exhibit 44 GHz oscillations with small amplitudes in the first 200 ps, which corresponds to the oscillation between the polaritonic eigenstates of the first rung of the dressed ladder and is analogous to the experimentally observed oscillations in single atom-cavity systems \cite{rempe}. The (green/dashed) polaritonic PB trace also oscillates at 7.7 GHz at longer times, whose origin we assign to the interference between the upper polaritonic and the subradiant states. This trend is also observed for other sets of parameters.

\begin{figure}[t]
\includegraphics[width=78mm]{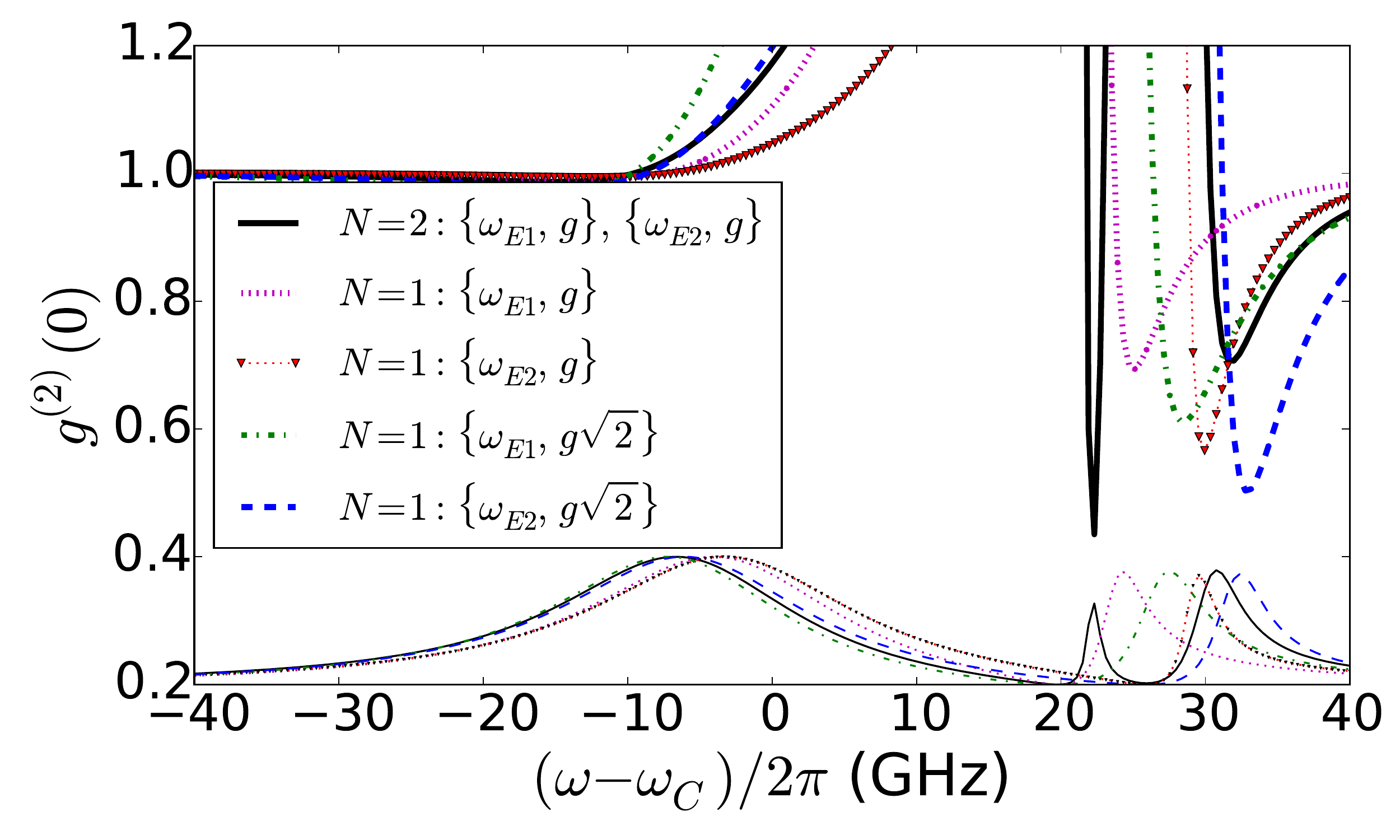}
\caption{\label{fig5} Comparison between second-order coherence in $N=2$ and $N=1$ emitter-cavity systems; emission spectra in thin lines at the bottom illustrate the relationship between features. Each emitter is characterized by frequency and coupling rate, $\{\omega_{E1}, \omega_{E2}, g\}/2\pi=\{20, 26, 10\}$ GHz.}
\end{figure}

\emph{Discussion}---To illustrate the advantages of multi-emitter over single-emitter cavity quantum electrodynamics we plot the second-order coherence as a function of laser detuning for comparable $N=2$ and $N=1$ systems (Fig. \ref{fig5}). First, we notice that, in addition to the polaritonic PB dip, only the two-emitter system (black solid line) exhibits the subradiant PB dip characterized in this paper. Next, its $g^{(2)}(0)$ value is lower than the one of the individually coupled emitters (magenta/dotted and red/triangles), and even of the individual emitters coupled with a collective rate for a two-emitter system $g\sqrt{2}$ (green/dashdot and blue/dashed). From a practical point of view, in addition to providing lower $g^{(2)}(0)$ value, transmission through the subradiant state extends an opportunity for a more robust single photon generation. The frequency and the second-order coherence values of light transmitted through the subradiant state are close to constant for variable emitter detuning. This implies that for any pair of nonidentical emitters (3 GHz $\leq \Delta_E/2\pi \leq$ 20 GHz) the quality of single photon emission is governed by the cavity detuning from the first emitter. This controllability is experimentally accessible as the cavity detuning can be controlled by gas tuning techniques without influencing the operating laser frequency \cite{alp}. Finally, the cross-polarized reflectivity method to addressing CQED systems \cite{dirk} can be applied to block the pump laser at the output channel.

In conclusion, we have analyzed nonclassical light generation in a strongly coupled two-emitter CQED system for variable cavity and emitter detuning. Combining quantum master equation and effective Hamiltonian approaches, we identified the parameters that give rise to a new, subradiant, mechanism of robust photon blockade, and explained their origin in the overlap between the eigenstates of multiple rungs of the dressed ladder. We also characterized the oscillations in $g^{(2)}(\tau)$ function as interference between the states of the first rung. The time scale of the single photon emission represents an order of magnitude speedup over the bare emitter dynamics, while the $g^{(2)}(0)$ values are improved over the system with a single emitter in a cavity. In light of the presented opportunities in polaritonic, subradiant and unconventional photon blockade, multi-photon emission and operating rate speedup, we expect that $N>2$ multi-emitter CQED systems will unveil even richer physics. For systems with more than several emitters, numerical calculations may prove lengthy due to the large size of the density matrix, however, our theoretical analysis based on the diagonalization of the effective Hamiltonian can provide a quick insight into the potential parameter areas with robust single photon emission. Finally, the introduction of dephasing into the model will help perform an even more accurate study of more realistic cavity quantum electrodynamics systems.

\emph{Acknowledgements}---This material is based upon work supported by the Air Force Office of Scientific Research under award number FA9550-17-1-0002 and by the National Science Foundation (DMR Grant Numbers 1406028 and 1503759). We thank Hideo Mabuchi, Nikolas Tezak and Dmitri Pavlichin for constructive discussions. K.A.F. acknowledges support from the Lu Stanford Graduate Fellowship and the National Defense Science and Engineering Graduate Fellowship.

\bibliography{two-emitter}% Produces the bibliography via BibTeX.

\end{document}